\documentclass[twocolumn,aps,pra,showpacs,floatfix,superscriptaddress]{revtex4-1}
\usepackage{graphicx}
\usepackage{amsmath}
\usepackage{bm}
\usepackage{color}
\usepackage{dcolumn}
\usepackage{hyperref}

\begin{document}


\newcommand{\etal}{{\it et al.}}
\newcommand{\fref}[1]{Fig.~\ref{#1}}
\newcommand{\Fref}[1]{Figure \ref{#1}}
\newcommand{\sref}[1]{Sec.~\ref{#1}}
\newcommand{\Eref}[1]{Eq.~(\ref{#1})}
\newcommand{\tref}[1]{Table~\ref{#1}}
\newcommand{\rtw}{\longrightarrow}
\def\veps{\varepsilon}
\newcommand{\cm}{\ensuremath{\mathrm{cm}^{-1}}}

\newcommand{\cmt}[1]{[\![#1]\!]}
\newcommand{\au}{\ensuremath{\mathrm{a.u.}}}
\newcommand{\pprime}{{\prime\prime}}
\newcommand{\g}{\ensuremath{{^1\Sigma_g^+}}}
\newcommand{\e}{\ensuremath{{^3\Sigma_u^+}}}
\newcommand{\gx}{\ensuremath{X\,{^1\Sigma_g^+}}}
\newcommand{\ex}{\ensuremath{a\,{^3\Sigma_u^+}}}
\newcommand{\dem}{DeMille \etal}
\newcommand{\zel}{Zelevinsky \etal}

\newcommand{\NZIAS}{
Centre for Theoretical Chemistry and Physics,
New Zealand Institute for Advanced Study,
Massey University, Auckland 0745, New Zealand}

\newcommand{\UNSW}{
School of Physics, University of New South Wales, Sydney 2052, Australia}

\title{Effect of $\alpha$ variation on a prospective experiment to detect variation of $m_e/m_p$ in diatomic molecules}

\author{K. Beloy}
\affiliation{\NZIAS}

\author{A. Borschevsky}
\affiliation{\NZIAS}

\author{V. V. Flambaum}
\affiliation{\UNSW}
\affiliation{\NZIAS}

\author{P. Schwerdtfeger}
\affiliation{\NZIAS}

\date{\today}

\begin{abstract}
We consider the influence of variation in the fine structure constant $\alpha$ on a promising experiment proposed by \dem~to search for variation in the electron-to-proton mass ratio $\mu$ using diatomic molecules [\dem, Phys.~Rev.~Lett.~{\bf 100}, 043202 (2008)]. The proposed experiment involves spectroscopically probing the splitting between two nearly-degenerate vibrational levels supported by different electronic potentials. 
Here we demonstrate that this splitting may be equally or more sensitive to variation in $\alpha$ as to variation in $\mu$. For the anticipated experimental precision, this implies that the $\alpha$ variation may not be negligible, as previously assumed, and further suggests that the method could serve as a competitive means to search for $\alpha$ variation as well.
\end{abstract}


\maketitle


\section{Introduction}

It is conventionally assumed that the fine structure constant $\alpha\equiv e^2/\hbar c$ and electron-to-proton mass ratio $\mu\equiv m_e/m_p$ are non-varying physical quantities. With a successful track record dating back to the developments of quantum mechanics and atomic theory, the legitimacy of these assumptions is often taken for granted.
Modern experiments have verified the stability of these respective quantities on the fractional level of $10^{-17}$ and $10^{-14}$ per year~\cite{RosHumSch08etal,SheButCha08}. 
Still the search for variations in $\alpha$, $\mu$, or other fundamental ``constants'' has been motivated by theoretical attempts to unify gravity with the other fundamental forces of nature, with some leading models suggesting temporal or spatial dependence of these quantities \cite{Uza03}. Moreover, astrophysical data has been used to give evidence for nonzero variation of $\alpha$ and $\mu$ over cosmological time and distance scales~\cite{WebMurFla01etal,WebKinMur10etal,ReiBunHol06etal} (see also null results of Refs.~\cite{FlaKoz07b,SriChaPet04,HenMenMur09etal,Kan11,MulBeeGue10etal,AgaMolLev11}), further motivating efforts to detect signals of $\alpha$ and $\mu$ variation in the laboratory.

One promising method to search for variation of $\mu$ in diatomic molecules has been proposed by \dem~\cite{DeMSaiSag08etal}.
The envisioned experiment probes the splitting between two nearly-degenerate vibrational levels, with the two levels being supported by different electronic potentials. The authors show that for a given variation in $\mu$, the shift in transition frequency may be large on both an absolute scale and relative to the splitting itself. 
Further still, the authors experimentally identified a favorable transition in $^{133}$Cs$_2$, with one level associated with the ground \gx{} electronic potential and the other with the excited \ex{} electronic potential. The \gx{} and \ex{} potentials have a common dissociation limit, corresponding to two ground-state Cs atoms. Based on their Cs$_2$ analysis, the authors argue that fractional variations of $\mu$ on the level of $10^{-17}$ or smaller could be detected with their technique.

In this paper we consider the influence of $\alpha$ variation on this prospective experiment. Neglect of $\alpha$ variation is seemingly justified by the fact that the vibrational spectrum is independent of $\alpha$ in the nonrelativistic limit. Nevertheless, sensitivity arises from relativistic corrections to the electronic potential, 
provoking a closer examination of their effects on the proposed Cs$_2$ experiment. 
To facilitate in this analysis, we introduce coefficients quantifying the energy shift of individual levels in the vibrational spectrum for given fractional variations in $\mu$ and $\alpha$; we define these sensitivity coefficients according to the relation
\begin{eqnarray}
\delta E_v=
q_{\mu,v}\frac{\delta\mu}{\mu}
+q_{\alpha,v}\frac{\delta\alpha}{\alpha},
\label{Eq:varEv}
\end{eqnarray}
where $v$ labels the particular vibrational level. We employ atomic units (preceding arguments implicitly assumed this choice), though energies and sensitivity coefficients are often expressed numerically in \cm.
Furthermore, we choose the dissociation limit to be our zero of energy, with the convenience that this is common to both the \gx{} and \ex{} states. 
These specifications are required to unambiguously define the sensitivity coefficients of Eq.~(\ref{Eq:varEv}).

\section{Expressions for sensitivity coefficients}

Expressions for the sensitivity coefficients of Eq.~(\ref{Eq:varEv}) may be derived within the framework of the WKB approximation. For a given $\mu$, $\alpha$, and energy $E$, the phase is given by 
\begin{eqnarray}
\phi(\mu,\alpha,E)=
\int_{r_i(\alpha,E)}^{r_o(\alpha,E)}\sqrt{2M(\mu)\left[E-V(\alpha;r)\right]}dr,
\label{Eq:phi}
\end{eqnarray}
where $M(\mu)\propto\mu^{-1}$ is the reduced mass, $V(\alpha;r)$ is the electronic potential, which depends on the fine-structure constant in addition to the internuclear separation $r$, and $r_i(\alpha,E)$ and $r_o(\alpha,E)$ are the classical inner and outer turning points for energy $E$ (for brevity, we will subsequently refrain from writing dependence on $\mu$, $\alpha$, and $E$ explicitly).
Variations in $\mu$, $\alpha$, and $E$ impart a variation in $\phi$ given by
\begin{eqnarray}
\delta\phi=
\left(\frac{\partial\phi}{\partial\mu}\right)\delta\mu
+\left(\frac{\partial\phi}{\partial\alpha}\right)\delta\alpha
+\left(\frac{\partial\phi}{\partial E}\right)\delta E,
\label{Eq:varphi}
\end{eqnarray}
with the partial derivatives being
\begin{eqnarray}
\frac{\partial\phi}{\partial\mu}&=&-\frac{1}{2\mu}\phi,
\label{Eq:partialphimu}\\
\frac{\partial\phi}{\partial \alpha}&=&
-\int_{r_i}^{r_o}\sqrt{\frac{M}{2\left[E-V(r)\right]}}
\left(\frac{\partial V(r)}{\partial\alpha}\right)dr,
\nonumber\\
\frac{\partial\phi}{\partial E}&=&
\int_{r_i}^{r_o}\sqrt{\frac{M}{2\left[E-V(r)\right]}}dr.
\nonumber
\end{eqnarray}
Our choice for the zero of energy ensures both $V(r)=0$ and $\partial V(r)/\partial\alpha=0$ as $r\rightarrow\infty$.

Energy appears in Eq.~(\ref{Eq:phi}) as a continuous variable.
Boundary conditions imposed on the vibrational wave function restrict the values of physically allowed energies, and this may be accounted for by the WKB quantization condition:
\begin{eqnarray}
\phi_v=
\left(v+{\textstyle\frac{1}{2}}\right)\pi,
\label{Eq:WKBquantize}
\end{eqnarray}
with the energy satisfying this condition corresponding to the vibrational energy level $E_v$ (here $v$, the vibrational quantum number, is a non-negative integer). Eq.~(\ref{Eq:WKBquantize}) implies $\delta\phi_v=0$, which combined with Eq.~(\ref{Eq:varphi}) further gives
\begin{eqnarray}
\delta E_v=
-\frac{\left(\partial\phi/\partial\mu\right)_v}{\left(\partial\phi/\partial E\right)_v}\delta\mu
-\frac{\left(\partial\phi/\partial\alpha\right)_v}{\left(\partial\phi/\partial E\right)_v}\delta\alpha,
\label{Eq:varEv2}
\end{eqnarray}
where the subscripts $v$ on the right-hand-side denote evaluation at energy $E_v$ (evaluation at present-day values of $\alpha$ and $\mu$ is further implied). 
Taking the definitions
\begin{eqnarray}
q_{\mu}&\equiv&
\frac{\int_{r_i}^{r_o}\left[E-V(r)\right]^{+1/2}dr}
{\int_{r_i}^{r_o}\left[E-V(r)\right]^{-1/2}dr},
\label{Eq:qmu}
\\
q_{\alpha}&\equiv&
\frac{\int_{r_i}^{r_o}\left[E-V(r)\right]^{-1/2}
\left(\alpha\frac{\partial V(r)}{\partial\alpha}\right)dr}
{\int_{r_i}^{r_o}\left[E-V(r)\right]^{-1/2}dr},
\label{Eq:qalpha}
\end{eqnarray}
the sensitivity coefficients $q_{\mu,v}$ and $q_{\alpha,v}$ appearing in Eq.~(\ref{Eq:varEv}) may then be associated with $q_{\mu}$ and $q_{\alpha}$ evaluated at energy $E_v$. We choose to use the term sensitivity factors to distinguish $q_{\mu}$ and $q_{\alpha}$ from the sensitivity coefficients $q_{\mu,v}$ and $q_{\alpha,v}$. It is worth noting that, although $E_v$ depends on the reduced mass and therefore the electron-to-proton mass ratio, the sensitivity factors themselves do not. We further note that $q_\mu$ is necessarily non-negative.

The WKB quantization condition (\ref{Eq:WKBquantize}) tells us that as we climb the energy spectrum, each new vibrational state is associated with an incremental change of $\pi$ in the phase $\phi$.
We may thus associate $\rho\equiv\pi^{-1}(\partial\phi/\partial E)$ with the density of states at energy $E$. It then follows from Eqs.~(\ref{Eq:partialphimu},\ref{Eq:WKBquantize},\ref{Eq:varEv2}) that the sensitivity coefficient $q_{\mu,v}$ may be simply expressed as
\begin{eqnarray}
q_{\mu,v}=\frac{(v+{\textstyle\frac{1}{2}})}{2\rho_v},
\label{Eq:qmuDeMille}
\end{eqnarray}
where $\rho_v$ is the density of states at energy $E_v$.

Expression (\ref{Eq:qmuDeMille}) was presented in the Letter of DeMille~\etal~\cite{DeMSaiSag08etal}. 
The authors noted that while the density of states is essentially constant for the lowest $v$ (the vibrational states being well-described by those of a harmonic oscillator), it rapidly increases for the highest $v$. Together with the numerator of Eq.~(\ref{Eq:qmuDeMille}), this suggests that maximum sensitivity to $\mu$ variation occurs within the intermediate part of the vibrational spectrum. This was a foundational principle for their proposal.
While Eq.~(\ref{Eq:qmuDeMille}) perhaps provides a more tangible means to visualize this behavior, Eq.~(\ref{Eq:qmu}) will be of greater operational use for us.

\section{Modeling $V(r)$ and $\partial V(r)/\partial\alpha$}

\begin{figure}[t]
\begin{center}
\includegraphics*[scale=0.47]{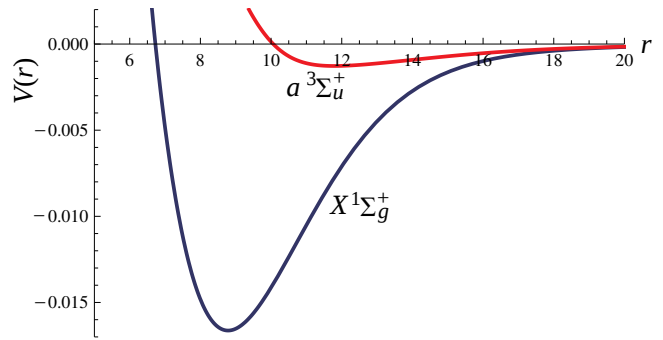}
\end{center}
\caption{(color online) 
Potential energy curves for the \gx{} and \ex{} states of Cs$_2$ as given by analytical functions of Ref.~\cite{CoxHaj10} and Ref.~\cite{XieSovLyy09etal}, respectively.
All units are atomic units.}
\label{Fig:Cs2forPeter}
\end{figure}

Coxon and Hajigeorgiou \cite{CoxHaj10} and Xie \etal~\cite{XieSovLyy09etal} have presented analytical potential energy curves based on accurate fits to experimental data of the \gx{} and \ex{} states of Cs$_2$, respectively. These are illustrated in Fig.~\ref{Fig:Cs2forPeter}. With these curves, we may determine $q_\mu$ for both states directly via Eq.~(\ref{Eq:qmu}). To further determine $q_\alpha$ [Eq.~(\ref{Eq:qalpha})], we must also know the change in the potentials $V(r)$ with respect to a change in $\alpha$;
this we model from computed data.

We begin by describing our determination of $\partial V(r)/\partial\alpha$ for the ground state. We have previously computed the potential energy curve for this state in considering $\alpha$ variation in ultracold atomic collision experiments \cite{BorBelFla11}; we employ this data for our present purposes as well. This data was obtained using the relativistic DIRAC computational program \cite{DIRAC10} within a coupled cluster singles-doubles (CCSD) approximation (see Ref.~\cite{BorBelFla11} for further details).
Our data agrees very well with the analytical curve in the vicinity of the equilibrium distance and at shorter distances, but fails to produce the appropriate assymptotic behavior, $-C_6/r^6$, ultimately resulting in a well-depth 10\% too large (see Fig.~1 in Ref.~\cite{BorBelFla11}).
The divergence of our data from the correct asymptotic curve is due in large, we suspect, to basis set limitations and neglect of higher excitations (triples, quadruples, etc.)~in the CCSD approximation.

\begin{figure}[t]
\begin{center}
\includegraphics*[scale=0.42]{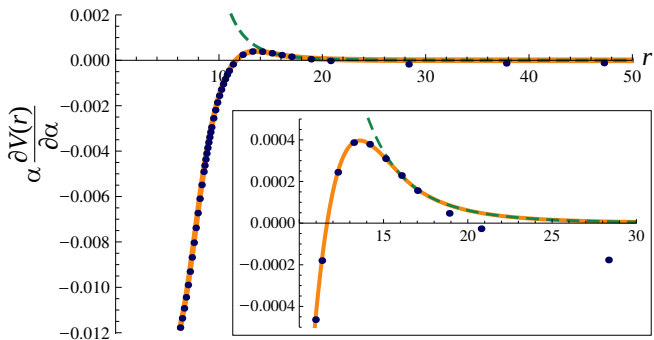}
\end{center}
\caption{(color online) 
Change in the potential energy curve of the Cs$_2$ ground state with respect to fractional change in $\alpha$.
The circles correspond to computed data. The dashed line represents the leading asymptotic term, $-\alpha(\partial C_6/\partial\alpha)/r^6$, with $\partial C_6/\partial\alpha$ as estimated in Ref.~\cite{BorBelFla11}. The solid line is a fit to the calculated data points in the shorter range with a smooth transition to the asymptotic curve. The inset magnifies the region of the maximum.
All units are atomic units.}
\label{Fig:Cs2changecurve}
\end{figure}

The potential energy curve was computed with various values of $\alpha$ in the neighborhood of $\alpha=1/137$. With numerical differentiation with respect to $\alpha$, we obtain $\partial V(r)/\partial\alpha$. Our computed data is displayed in Fig.~\ref{Fig:Cs2changecurve}. In principle, an offset for these data points should be chosen to fulfill the criteria $\partial V(r)/\partial\alpha=0$ as $r\rightarrow\infty$. However, as our computed data fails to produce the correct asymptotic behavior for $V(r)$, it is not expected to produce the correct asymptotic behavior for $\partial V(r)/\partial\alpha$ either. For a more appropriate representation, we choose to model the asymptotic part by $-(\partial C_6/\partial\alpha)/r^6$, using the estimate of $\partial C_6/\partial\alpha$ given in Ref.~\cite{BorBelFla11}. We join this asymptotic curve smoothly with a fit of our computed data points in the shorter range to obtain $\partial V(r)/\partial\alpha$; the resulting curve is illustrated in Fig.~\ref{Fig:Cs2changecurve}. The fractional change in potential depth $D$ with respect to fractional change in $\alpha$ is found to be $\partial\mathrm{ln}(D)/\partial\mathrm{ln}(\alpha)=0.25$.

A notable feature of $\partial V(r)/\partial\alpha$ is its non-monotonic behavior. This is expected based on two qualitative features of our computed data. Firstly, the potential is found to get deeper for an increase in $\alpha$; correspondingly, $\partial V(r)/\partial\alpha$ is negative in the vicinity of $r_e$. Secondly, the $C_6$ coefficient becomes smaller for an increase in $\alpha$, a result which may be related to the fact that relativistic effects diminish the ground state polarizability of atomic Cs~\cite{LimPerSet99etal,LimSchMet05,BorBelFla11}. The implication is a positive-valued $\partial V(r)/\partial\alpha$ in the asymptotic region. Inevitably $\partial V(r)/\partial\alpha$ must have a maximum in the intermediate region, which indeed appears in our computed data. The non-monotonic behavior of $\partial V(r)/\partial\alpha$ has implications for $q_\alpha$ which will be discussed below. We also note that $\partial V(r)/\partial\alpha$ is essentially linear in the vicinity of $r_e$ and at shorter distances.

The \ex{} potential energy curve proves more difficult to compute accurately than the ground state, and so we are led to model $\partial V(r)/\partial\alpha$ for this state in a less sophisticated manner. From a relativistic Fock space coupled-cluster calculation, we estimate the fractional change in potential depth $D$ with respect to fractional change in $\alpha$ to be $\partial\mathrm{ln}(D)/\partial\mathrm{ln}(\alpha)\approx-0.17$. This is comparable in magnitude to the value obtained for the ground state, but with an opposite sign. The negative sign here implies that the potential becomes shallower for an increase in $\alpha$. To model $\partial V(r)/\partial\alpha$, we begin by assuming linear behavior in the vicinity of the equilibrium distance and at shorter range, taking a fixed value of $0.17(D/\alpha)$ at $r_e$. We smoothly join this to the same asymptotic curve as the \gx{} state, with the reasoning that the $C_6$ coefficient is common to both states. Qualitatively, we expect this to be an accurate depiction of $\partial V(r)/\partial\alpha$. Namely, this model predicts $\partial V(r)/\partial\alpha$ to be positive and monotonically decreasing with $r$.

\section{Estimated sensitivity factors}

\begin{figure*}[t]
\begin{center}
\includegraphics*[scale=0.42]{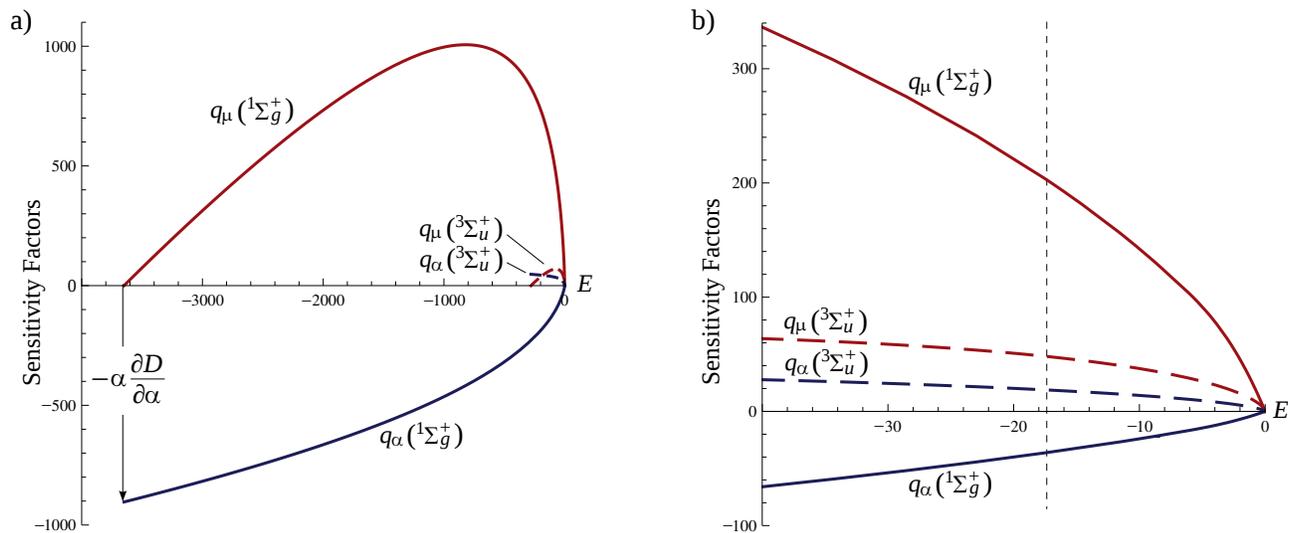}
\end{center}
\caption{(color online) Sensitivity factors $q_\mu$ and $q_\alpha$ for the \gx{} and \ex{} states of Cs$_2$ versus energy $E$. Each curve extends from $E=0$ to $E=-D$, where $D$ is the potential depth of the respective state. 
Panel (a) displays the full extent of the curves, while (b) provides a magnification of the low-$|E|$ region. The vertical dashed line in (b) identifies the approximate location of nearly-degenerate vibrational levels (for $^{133}$Cs$_2$), with one level being supported by the ground state potential and the other being supported by the excited state potential \cite{DeMSaiSag08etal}.
All values are expressed in \cm.}
\label{Fig:qalphaqmu}
\end{figure*}

With $V(r)$ from Refs.~\cite{CoxHaj10,XieSovLyy09etal} and $\partial V(r)/\partial\alpha$ modeled from computed data, we have the tools required to evaluate the sensitivity factors $q_\mu$ and $q_\alpha$ via Eqs.~(\ref{Eq:qmu},\ref{Eq:qalpha}). These are displayed for both the \gx{} and \ex{} states in Fig.~\ref{Fig:qalphaqmu}(a) as a function of energy $E$. The first thing to note is that, generally speaking, $q_\mu$ and $q_\alpha$ are of similar magnitude for the respective states, with a maximum absolute value in the range of (0.17---0.28)$D$ in each case. However, while $q_\mu$ peaks at $E\approx-D/4$ for both states, $q_\alpha$ peaks in absolute value at the bottom of the potential well, $E=-D$.

The absolute shift to a transition frequency 
is determined by the difference in sensitivity coefficients for the two levels involved in the transition. \dem~suggested probing the transition between two nearly degenerate vibrational levels in Cs$_2$, with one level supported by the \gx{} state and the other supported by the \ex{} state. The larger scale of $q_\mu$ for the \gx{} state compared to the \ex{} state ensures a significant difference in sensitivity coefficients, provided sufficiently bound levels are chosen. 
The specification of nearly degenerate levels is motivated by the fact that, in addition to large absolute sensitivity of the transition, experimental considerations further call for large sensitivity relative to the splitting itself \cite{DeMSaiSag08etal}; a small splitting translates to a large relative sensitivity.

For the described experiment, absolute sensitivity of the splitting $\omega$ to variations in $\alpha$ and $\mu$ may then be gauged by $\Delta q_\mu$ and $\Delta q_\alpha$---the differences between sensitivity factors of the \gx{} and \ex{} states---evaluated at the energy of the near degeneracy. From inspection of Fig.~\ref{Fig:qalphaqmu}(a), we see that $|\Delta q_\mu|$ and $|\Delta q_\alpha|$ are both maximum at the bottom of the \ex{} potential well, $E=-279~\cm$ ($\Delta q_\mu$ and $\Delta q_\alpha$ lose meaning for energies below this). From the perspective of absolute sensitivity to $\mu$ variation, a transition in this region would be the most favorable to probe. However, here the density of states is smallest for the two states, making near degeneracies less likely to occur. Approaching the dissociation limit ($E=0$), the densities rapidly increase and a suitably small $\omega$ is more likely. \dem~have experimentally identified one transition in $^{133}$Cs$_2$ suitable for their proposed method, the near degeneracy occurring at approximately $E=-17.4~\cm$
\cite{DeMSaiSag08etal}. Indeed, these levels are high in the vibrational spectrum, having binding energy 0.5\% and 6\% of the potential depth for the \gx{}  and \ex{} states, respectively. The actual value of $\omega$ here depends on the particular transition selected from the hyperfine-rotational substructure, \dem~having suggested two possibilities with $\omega\sim0.1~\cm$~\cite{DeMSaiSag08etal}.

Fig.~\ref{Fig:qalphaqmu}(b) provides a magnification of the sensitivity factors in the low-$|E|$ region, including the location of the near degeneracy identified by \dem{} In this region, we see that the \ex{} state has a rather sizable contribution to $\Delta q_\mu$ and $\Delta q_\alpha$. Moreover, as $q_\mu$ is positive for both states, these sensitivity factors cancel to some degree in $\Delta q_\mu$ (i.e., the levels move in the same direction with respect to variation in $\mu$). On the other hand, the $q_\alpha$ are seen to have opposite signs for the two states, thus having a constructive effect in $\Delta q_\alpha$ (i.e., the levels move in the opposite directions with respect to variation in $\alpha$). This may be attributed, in large, to the fact that a variation in $\alpha$ causes one potential to get deeper and the other to get shallower.

In Table~\ref{Tab:rhoandqs} we present values for the sensitivity factors $q_\mu$ and $q_\alpha$ and differential sensitivity factors $\Delta q_\mu$ and $\Delta q_\alpha$ evaluated at the energies $E=-279~\cm$ and $E=-17.4~\cm$, corresponding to the bottom of the \ex{} potential and the location of the near degeneracy identified by \dem{} 
For both energies we see that $|\Delta q_\alpha|$ is only a factor of $\sim\!3$ less than $|\Delta q_\mu|$. 
This ratio is found hold for intermediate energies as well.
Thus we write the following approximate relation for the shift in $\omega$ with respect to variations in $\mu$ and $\alpha$:
\begin{equation*}
\delta\omega\approx K\left(\frac{\delta\mu}{\mu}-\frac{1}{3}\frac{\delta\alpha}{\alpha}\right).
\end{equation*}
Specifically for the near degeneracy identified by \dem, $|K|\approx160~\cm$.

\begin{table}[t]
\caption{Sensitivity factors for the \gx{} and \ex{} states of $^{133}$Cs$_2$ evaluated at $E=-279~\cm$ and $E=-17.4~\cm$, these energies corresponding to the bottom of the \ex{} potential and the approximate location of a near-degeneracy between vibrational levels \cite{DeMSaiSag08etal}, respectively. $\Delta$ indicates the difference between \gx{} and \ex{} state values. The density of states $\rho$ is also provided in each case.}
\label{Tab:rhoandqs}
\begin{center}
\begin{ruledtabular}
\begin{tabular}{lD{.}{.}{3.0}D{.}{.}{3.0}D{.}{.}{3.0}}%
& \multicolumn{1}{c}{\gx} 
& \multicolumn{1}{c}{\ex} 
& \multicolumn{1}{c}{$\Delta$} 
\\
\hline
\\[-2mm]
&\multicolumn{3}{c}{Evaluated at $E=-279~\cm$}\\
$q_\mu~[\cm]$ &  827 &  0 & 827 \\
$q_\alpha~[\cm]$ & -229 & 47 & -276 \\
$\rho~[1/\cm]$ & \multicolumn{1}{c}{0.065} & \multicolumn{1}{c}{0.086} & \\
\\[-2mm]
&\multicolumn{3}{c}{Evaluated at $E=-17.4~\cm$}\\
$q_\mu~[\cm]$ &  203 &  48 & 155 \\
$q_\alpha~[\cm]$ & -36 & 19 & -55 \\
$\rho~[1/\cm]$ & \multicolumn{1}{c}{0.34} & \multicolumn{1}{c}{0.39} & \\
\end{tabular}
\end{ruledtabular}
\end{center}
\end{table}

In practice, the transition frequency must be measured with respect to some reference (clock) frequency $\omega_c$, and only variation in the ratio $\Omega\equiv\omega/\omega_c$ may be extracted. (We reiterate here that $\omega$ and $\omega_c$ represent the transition and clock frequencies given in atomic units; $\Omega$ is effectively the transition frequency in units of the clock frequency.) Evidently the shift in the clock frequency must be further taken into account according to the relation
\begin{equation*}
\frac{\delta\Omega}{\Omega}=\frac{\delta\omega}{\omega}-\frac{\delta\omega_c}{\omega_c}.
\end{equation*}
\dem~suggest using an optical atomic clock as a reference, in which case $\omega_c$ is essentially independent of $\mu$. $\alpha$ dependence has been considered for species currently used as optical standards \cite{DzuFlaWeb99a,AngDzuFla04}. For Hg$^+$, $|\delta\omega_c/\omega_c|\approx2|\delta\alpha/\alpha|$, while other clocks are much less sensitive to $\alpha$ variation. Thus for $|K|/\omega\gg1$, as with the transition of interest in $^{133}$Cs$_2$, the shift in $\omega_c$ need not be considered.

Before concluding, we briefly discuss the accuracy of our estimates, focusing on the results for $E=-17.4~\cm$. For the \gx{} potential, the classical inner and outer turning points are found to be 6.7~\au~and 22~\au~using the analytical potential of Ref.~\cite{CoxHaj10}. From Fig.~\ref{Fig:Cs2changecurve}, we see that $\partial V(r)/\partial\alpha$ has both negative and positive valued segments over this range. We may decompose $q_\alpha$ into negative and positive contributions accordingly, 
and in doing so we find
\begin{equation*}
q_\alpha=-54~\cm+18~\cm,
\end{equation*}
where the two terms represent the respective signed contributions. We see that there is a large degree of cancellation between these contributions, resulting in the value $q_\alpha=-36~\cm$ found in Table~\ref{Tab:rhoandqs}. We tested the stability of our result using various models to match our data with the asymptotic form of $\partial V(r)/\partial\alpha$ (e.g., including order-of-magnitude estimates for variations in $C_8$ and $C_{10}$ coefficients). We find the negative contribution to be highly stable with respect to these different models. The positive contribution, on the other hand, is found to be quite dependent on the overall offset of our data points, this being determined by the particular model (see, e.g., Fig.~\ref{Fig:Cs2changecurve}). For the alternative models, the positive contribution never fluctuated by more than a factor of two.

For the \ex{} state, we predict $\partial V(r)/\partial\alpha$ to be positive and to decrease monotonically with $r$. Here there is no cancellation between contributions of different sign as with the \gx{} state. This, along with the smaller scale set by the potential depth, largely validates our less sophisticated modeling of $\partial V(r)/\partial\alpha$ in this case.

For a rough error estimate, we ascribe 100\% uncertainty to the positive contribution of our $q_\alpha$ for the \gx{} state and 100\% uncertainty to our value of $q_\alpha$ for the \ex{} state. We subsequently conclude that, at the location of the near degeneracy identified by \dem, $\Delta q_\alpha/\Delta q_\mu=-0.35$ with about 50\% uncertainty. This accuracy is sufficient to draw important qualitative conclusions in the following section.

Finally, we may compare our result with a cursory estimate found in the review of Flambaum and Kozlov \cite{FlaKoz08}. Here the authors predicted a somewhat weaker influence of $\alpha$ variation on the proposed Cs$_2$ experiment, effectively finding the ratio $\Delta q_\alpha/\Delta q_\mu$ to be more than a factor of two smaller than our present value (the authors also guessed the sign to be opposite). Their rudimentary estimate used atomic data in place of unkown molecular data and neglected important anharmonic effects of the potential; our present value is based on a more refined method.

\section{Conclusion}

Here we have considered the effect of $\alpha$ variation on a prospective experiment to search for variation of $\mu$ using nearly degenerate vibrational levels in $^{133}$Cs$_2$~\cite{DeMSaiSag08etal}. We estimate this experiment to be only a factor of three less sensitive to $\alpha$ variation as it is to $\mu$ variation.
In Ref.~\cite{DeMSaiSag08etal}, \dem~argued that this experiment could plausibly detect variation of $\mu$ at a fractional level of $10^{-17}$ or less. Our result shows that attaining experimental precision sensitive to fractional variation of $\mu$ at $1\times10^{-17}$, for example, implies an accompanying sensitivity to fractional variation of $\alpha$ at $3\times10^{-17}$. 
The most stringent laboratory limits to-date allow for annual drift of $\alpha$ at this level \cite{RosHumSch08etal}. Therefore, we conclude that $\alpha$ variation may not be negligible for the proposed experiment.

We can extend further on this conclusion by noting that Cs$_2$ was presented in Ref.~\cite{DeMSaiSag08etal} as a candidate system for a more general experimental method. 
Ultimately other diatomic systems may prove more advantageous, and theoretical work is currently underway with the goal of determining optimal systems for this method \cite{Kozlovprivate}. Other systems---more specifically, select transitions in other systems---may also be significantly more or less sensitive to $\alpha$ variation than the Cs$_2$ case considered here.
For example, heavier systems will have a higher density of states compared to lighter systems with a similar electronic potential. As alluded to earlier, higher density of states is a favorable feature for this method as it increases the likelihood for near-degeneracies to occur between vibrational levels. At the same time, heavier systems are also known to have larger relativistic effects and, presumably, larger sensitivities to $\alpha$ variation.

As another example, we note the close relationship between the proposal of \dem~and a previous proposal of Flambaum and Kozlov \cite{FlaKoz07a}. Flambaum and Kozlov suggested using nearly degenerate vibrational levels associated with different fine structure components of an electronic multiplet, specifically focusing on levels near the bottom of the spectrum. While allowing for large relative sensitivity, this particular method does not promote enhanced absolute sensitivity. Nevertheless, it is rather straightforward to show that such an experiment is four times more sensitive to $\alpha$ variation than to $\mu$ variation, following from the simple $\alpha^2$ and $\mu^{1/2}$ scaling of fine structure and vibrational intervals, respectively. Evidently the general method of \dem~has potential to be more sensitive to $\alpha$ variation than to $\mu$ variation.

Finally, reinterpreting our above results, we suggest that the general method proposed by \dem~to probe variation in the electron-to-proton mass ratio in diatomic molecules may be an equally viable method to probe variation in the fine structure constant.

\section{Acknowledgements}
This work was supported by the Marsden Fund, administered by the Royal Society of New Zealand. VF further acknowledges support by the ARC. 


\begin{thebibliography}{24}%
\makeatletter
\providecommand \@ifxundefined [1]{%
 \@ifx{#1\undefined}
}%
\providecommand \@ifnum [1]{%
 \ifnum #1\expandafter \@firstoftwo
 \else \expandafter \@secondoftwo
 \fi
}%
\providecommand \@ifx [1]{%
 \ifx #1\expandafter \@firstoftwo
 \else \expandafter \@secondoftwo
 \fi
}%
\providecommand \natexlab [1]{#1}%
\providecommand \enquote  [1]{``#1''}%
\providecommand \bibnamefont  [1]{#1}%
\providecommand \bibfnamefont [1]{#1}%
\providecommand \citenamefont [1]{#1}%
\providecommand \href@noop [0]{\@secondoftwo}%
\providecommand \href [0]{\begingroup \@sanitize@url \@href}%
\providecommand \@href[1]{\@@startlink{#1}\@@href}%
\providecommand \@@href[1]{\endgroup#1\@@endlink}%
\providecommand \@sanitize@url [0]{\catcode `\\12\catcode `\$12\catcode
  `\&12\catcode `\#12\catcode `\^12\catcode `\_12\catcode `\%12\relax}%
\providecommand \@@startlink[1]{}%
\providecommand \@@endlink[0]{}%
\providecommand \url  [0]{\begingroup\@sanitize@url \@url }%
\providecommand \@url [1]{\endgroup\@href {#1}{\urlprefix }}%
\providecommand \urlprefix  [0]{URL }%
\providecommand \Eprint [0]{\href }%
\providecommand \doibase [0]{http://dx.doi.org/}%
\providecommand \selectlanguage [0]{\@gobble}%
\providecommand \bibinfo  [0]{\@secondoftwo}%
\providecommand \bibfield  [0]{\@secondoftwo}%
\providecommand \translation [1]{[#1]}%
\providecommand \BibitemOpen [0]{}%
\providecommand \bibitemStop [0]{}%
\providecommand \bibitemNoStop [0]{.\EOS\space}%
\providecommand \EOS [0]{\spacefactor3000\relax}%
\providecommand \BibitemShut  [1]{\csname bibitem#1\endcsname}%
\let\auto@bib@innerbib\@empty
\bibitem [{\citenamefont {Rosenband}\ \emph {et~al.}(2008)\citenamefont
  {Rosenband} \emph {et~al.}}]{RosHumSch08etal}%
  \BibitemOpen
  \bibfield  {author} {\bibinfo {author} {\bibfnamefont {T.}~\bibnamefont
  {Rosenband}} \emph {et~al.},\ }\href {\doibase 10.1126/science.1154622}
  {\bibfield  {journal} {\bibinfo  {journal} {Science}\ }\textbf {\bibinfo
  {volume} {319}},\ \bibinfo {pages} {1808} (\bibinfo {year}
  {2008})}\BibitemShut {NoStop}%
\bibitem [{\citenamefont {Shelkovnikov}\ \emph {et~al.}(2008)\citenamefont
  {Shelkovnikov}, \citenamefont {Butcher}, \citenamefont {Chardonnet},\ and\
  \citenamefont {Amy-Klein}}]{SheButCha08}%
  \BibitemOpen
  \bibfield  {author} {\bibinfo {author} {\bibfnamefont {A.}~\bibnamefont
  {Shelkovnikov}}, \bibinfo {author} {\bibfnamefont {R.~J.}\ \bibnamefont
  {Butcher}}, \bibinfo {author} {\bibfnamefont {C.}~\bibnamefont {Chardonnet}},
  \ and\ \bibinfo {author} {\bibfnamefont {A.}~\bibnamefont {Amy-Klein}},\
  }\href {\doibase 10.1103/PhysRevLett.100.150801} {\bibfield  {journal}
  {\bibinfo  {journal} {Phys. Rev. Lett.}\ }\textbf {\bibinfo {volume} {100}},\
  \bibinfo {pages} {150801} (\bibinfo {year} {2008})}\BibitemShut {NoStop}%
\bibitem [{\citenamefont {Uzan}(2003)}]{Uza03}%
  \BibitemOpen
  \bibfield  {author} {\bibinfo {author} {\bibfnamefont {J.-P.}\ \bibnamefont
  {Uzan}},\ }\href {\doibase 10.1103/RevModPhys.75.403} {\bibfield  {journal}
  {\bibinfo  {journal} {Rev. Mod. Phys.}\ }\textbf {\bibinfo {volume} {75}},\
  \bibinfo {pages} {403} (\bibinfo {year} {2003})}\BibitemShut {NoStop}%
\bibitem [{\citenamefont {Webb}\ \emph {et~al.}(2001)\citenamefont {Webb} \emph
  {et~al.}}]{WebMurFla01etal}%
  \BibitemOpen
  \bibfield  {author} {\bibinfo {author} {\bibfnamefont {J.~K.}\ \bibnamefont
  {Webb}} \emph {et~al.},\ }\href {\doibase 10.1103/PhysRevLett.87.091301}
  {\bibfield  {journal} {\bibinfo  {journal} {Phys. Rev. Lett.}\ }\textbf
  {\bibinfo {volume} {87}},\ \bibinfo {pages} {091301} (\bibinfo {year}
  {2001})}\BibitemShut {NoStop}%
\bibitem [{\citenamefont {Webb}\ \emph {et~al.}()\citenamefont {Webb} \emph
  {et~al.}}]{WebKinMur10etal}%
  \BibitemOpen
  \bibfield  {author} {\bibinfo {author} {\bibfnamefont {J.~K.}\ \bibnamefont
  {Webb}} \emph {et~al.},\ }\href@noop {} {}\bibinfo {note} {{e-print
  arXiv:1008.3907v1}}\BibitemShut {NoStop}%
\bibitem [{\citenamefont {Reinhold}\ \emph {et~al.}(2006)\citenamefont
  {Reinhold} \emph {et~al.}}]{ReiBunHol06etal}%
  \BibitemOpen
  \bibfield  {author} {\bibinfo {author} {\bibfnamefont {E.}~\bibnamefont
  {Reinhold}} \emph {et~al.},\ }\href {\doibase 10.1103/PhysRevLett.96.151101}
  {\bibfield  {journal} {\bibinfo  {journal} {Phys. Rev. Lett.}\ }\textbf
  {\bibinfo {volume} {96}},\ \bibinfo {pages} {151101} (\bibinfo {year}
  {2006})}\BibitemShut {NoStop}%
\bibitem [{\citenamefont {Flambaum}\ and\ \citenamefont
  {Kozlov}(2007{\natexlab{a}})}]{FlaKoz07b}%
  \BibitemOpen
  \bibfield  {author} {\bibinfo {author} {\bibfnamefont {V.~V.}\ \bibnamefont
  {Flambaum}}\ and\ \bibinfo {author} {\bibfnamefont {M.~G.}\ \bibnamefont
  {Kozlov}},\ }\href {\doibase 10.1103/PhysRevLett.98.240801} {\bibfield
  {journal} {\bibinfo  {journal} {Phys. Rev. Lett.}\ }\textbf {\bibinfo
  {volume} {98}},\ \bibinfo {pages} {240801} (\bibinfo {year}
  {2007}{\natexlab{a}})}\BibitemShut {NoStop}%
\bibitem [{\citenamefont {Srianand}\ \emph {et~al.}(2004)\citenamefont
  {Srianand}, \citenamefont {Chand}, \citenamefont {Petitjean},\ and\
  \citenamefont {Aracil}}]{SriChaPet04}%
  \BibitemOpen
  \bibfield  {author} {\bibinfo {author} {\bibfnamefont {R.}~\bibnamefont
  {Srianand}}, \bibinfo {author} {\bibfnamefont {H.}~\bibnamefont {Chand}},
  \bibinfo {author} {\bibfnamefont {P.}~\bibnamefont {Petitjean}}, \ and\
  \bibinfo {author} {\bibfnamefont {B.}~\bibnamefont {Aracil}},\ }\href
  {\doibase 10.1103/PhysRevLett.92.121302} {\bibfield  {journal} {\bibinfo
  {journal} {Phys. Rev. Lett.}\ }\textbf {\bibinfo {volume} {92}},\ \bibinfo
  {pages} {121302} (\bibinfo {year} {2004})}\BibitemShut {NoStop}%
\bibitem [{\citenamefont {Henkel}\ \emph {et~al.}(2009)\citenamefont {Henkel}
  \emph {et~al.}}]{HenMenMur09etal}%
  \BibitemOpen
  \bibfield  {author} {\bibinfo {author} {\bibfnamefont {C.}~\bibnamefont
  {Henkel}} \emph {et~al.},\ }\href {\doibase 10.1051/0004-6361/200811475}
  {\bibfield  {journal} {\bibinfo  {journal} {Astron. Astrophys.}\ }\textbf
  {\bibinfo {volume} {500}},\ \bibinfo {pages} {725} (\bibinfo {year}
  {2009})}\BibitemShut {NoStop}%
\bibitem [{\citenamefont {Kanekar}(2011)}]{Kan11}%
  \BibitemOpen
  \bibfield  {author} {\bibinfo {author} {\bibfnamefont {N.}~\bibnamefont
  {Kanekar}},\ }\href@noop {} {\bibfield  {journal} {\bibinfo  {journal}
  {Astrophys. J. Lett.}\ }\textbf {\bibinfo {volume} {728}},\ \bibinfo {pages}
  {L12} (\bibinfo {year} {2011})}\BibitemShut {NoStop}%
\bibitem [{\citenamefont {Muller}\ \emph {et~al.}()\citenamefont {Muller} \emph
  {et~al.}}]{MulBeeGue10etal}%
  \BibitemOpen
  \bibfield  {author} {\bibinfo {author} {\bibfnamefont {S.}~\bibnamefont
  {Muller}} \emph {et~al.},\ }\href@noop {} {}\bibinfo {note} {{e-print
  arXiv:1104.3361v1}}\BibitemShut {NoStop}%
\bibitem [{\citenamefont {Agafonova}\ \emph {et~al.}(2011)\citenamefont
  {Agafonova}, \citenamefont {Molaro}, \citenamefont {Levshakov},\ and\
  \citenamefont {Hou}}]{AgaMolLev11}%
  \BibitemOpen
  \bibfield  {author} {\bibinfo {author} {\bibfnamefont {I.~I.}\ \bibnamefont
  {Agafonova}}, \bibinfo {author} {\bibfnamefont {P.}~\bibnamefont {Molaro}},
  \bibinfo {author} {\bibfnamefont {S.~A.}\ \bibnamefont {Levshakov}}, \ and\
  \bibinfo {author} {\bibfnamefont {J.~L.}\ \bibnamefont {Hou}},\ }\href@noop
  {} {\bibfield  {journal} {\bibinfo  {journal} {Astron. Astrophys.}\ }\textbf
  {\bibinfo {volume} {529}},\ \bibinfo {pages} {A28} (\bibinfo {year}
  {2011})}\BibitemShut {NoStop}%
\bibitem [{\citenamefont {DeMille}\ \emph {et~al.}(2008)\citenamefont {DeMille}
  \emph {et~al.}}]{DeMSaiSag08etal}%
  \BibitemOpen
  \bibfield  {author} {\bibinfo {author} {\bibfnamefont {D.}~\bibnamefont
  {DeMille}} \emph {et~al.},\ }\href {\doibase 10.1103/PhysRevLett.100.043202}
  {\bibfield  {journal} {\bibinfo  {journal} {Phys. Rev. Lett.}\ }\textbf
  {\bibinfo {volume} {100}},\ \bibinfo {pages} {043202} (\bibinfo {year}
  {2008})}\BibitemShut {NoStop}%
\bibitem [{\citenamefont {Coxon}\ and\ \citenamefont
  {Hajigeorgiou}(2010)}]{CoxHaj10}%
  \BibitemOpen
  \bibfield  {author} {\bibinfo {author} {\bibfnamefont {J.~A.}\ \bibnamefont
  {Coxon}}\ and\ \bibinfo {author} {\bibfnamefont {P.~G.}\ \bibnamefont
  {Hajigeorgiou}},\ }\href@noop {} {\bibfield  {journal} {\bibinfo  {journal}
  {J. Chem. Phys.}\ }\textbf {\bibinfo {volume} {132}},\ \bibinfo {pages}
  {094105} (\bibinfo {year} {2010})}\BibitemShut {NoStop}%
\bibitem [{\citenamefont {Xie}\ \emph {et~al.}(2009)\citenamefont {Xie} \emph
  {et~al.}}]{XieSovLyy09etal}%
  \BibitemOpen
  \bibfield  {author} {\bibinfo {author} {\bibfnamefont {F.}~\bibnamefont
  {Xie}} \emph {et~al.},\ }\href@noop {} {\bibfield  {journal} {\bibinfo
  {journal} {J. Chem. Phys.}\ }\textbf {\bibinfo {volume} {130}},\ \bibinfo
  {pages} {051102} (\bibinfo {year} {2009})}\BibitemShut {NoStop}%
\bibitem [{\citenamefont {Borschevsky}\ \emph {et~al.}(2011)\citenamefont
  {Borschevsky}, \citenamefont {Beloy}, \citenamefont {Flambaum},\ and\
  \citenamefont {Schwerdtfeger}}]{BorBelFla11}%
  \BibitemOpen
  \bibfield  {author} {\bibinfo {author} {\bibfnamefont {A.}~\bibnamefont
  {Borschevsky}}, \bibinfo {author} {\bibfnamefont {K.}~\bibnamefont {Beloy}},
  \bibinfo {author} {\bibfnamefont {V.~V.}\ \bibnamefont {Flambaum}}, \ and\
  \bibinfo {author} {\bibfnamefont {P.}~\bibnamefont {Schwerdtfeger}},\ }\href
  {\doibase 10.1103/PhysRevA.83.052706} {\bibfield  {journal} {\bibinfo
  {journal} {Phys. Rev. A}\ }\textbf {\bibinfo {volume} {83}},\ \bibinfo
  {pages} {052706} (\bibinfo {year} {2011})}\BibitemShut {NoStop}%
\bibitem [{\citenamefont {Saue}\ \emph {et~al.}()\citenamefont {Saue} \emph
  {et~al.}}]{DIRAC10}%
  \BibitemOpen
  \bibfield  {author} {\bibinfo {author} {\bibfnamefont {T.}~\bibnamefont
  {Saue}} \emph {et~al.},\ }\href@noop {} {}\bibinfo {note} {{{\it DIRAC, a
  relativistic {\it ab initio} electronic structure program, release DIRAC10}
  (see http:/\!/dirac.chem.vu.nl)}}\BibitemShut {NoStop}%
\bibitem [{\citenamefont {Lim}\ \emph {et~al.}(1999)\citenamefont {Lim} \emph
  {et~al.}}]{LimPerSet99etal}%
  \BibitemOpen
  \bibfield  {author} {\bibinfo {author} {\bibfnamefont {I.~S.}\ \bibnamefont
  {Lim}} \emph {et~al.},\ }\href {\doibase 10.1103/PhysRevA.60.2822} {\bibfield
   {journal} {\bibinfo  {journal} {Phys. Rev. A}\ }\textbf {\bibinfo {volume}
  {60}},\ \bibinfo {pages} {2822} (\bibinfo {year} {1999})}\BibitemShut
  {NoStop}%
\bibitem [{\citenamefont {Lim}\ \emph {et~al.}(2005)\citenamefont {Lim},
  \citenamefont {Schwerdtfeger}, \citenamefont {Metz},\ and\ \citenamefont
  {Stoll}}]{LimSchMet05}%
  \BibitemOpen
  \bibfield  {author} {\bibinfo {author} {\bibfnamefont {I.~S.}\ \bibnamefont
  {Lim}}, \bibinfo {author} {\bibfnamefont {P.}~\bibnamefont {Schwerdtfeger}},
  \bibinfo {author} {\bibfnamefont {B.}~\bibnamefont {Metz}}, \ and\ \bibinfo
  {author} {\bibfnamefont {H.}~\bibnamefont {Stoll}},\ }\href {\doibase
  DOI:10.1063/1.1856451} {\ \textbf {\bibinfo {volume} {122}},\ \bibinfo
  {pages} {104103} (\bibinfo {year} {2005})}\BibitemShut {NoStop}%
\bibitem [{\citenamefont {Dzuba}\ \emph {et~al.}(1999)\citenamefont {Dzuba},
  \citenamefont {Flambaum},\ and\ \citenamefont {Webb}}]{DzuFlaWeb99a}%
  \BibitemOpen
  \bibfield  {author} {\bibinfo {author} {\bibfnamefont {V.~A.}\ \bibnamefont
  {Dzuba}}, \bibinfo {author} {\bibfnamefont {V.~V.}\ \bibnamefont {Flambaum}},
  \ and\ \bibinfo {author} {\bibfnamefont {J.~K.}\ \bibnamefont {Webb}},\
  }\href {\doibase 10.1103/PhysRevA.59.230} {\bibfield  {journal} {\bibinfo
  {journal} {Phys. Rev. A}\ }\textbf {\bibinfo {volume} {59}},\ \bibinfo
  {pages} {230} (\bibinfo {year} {1999})}\BibitemShut {NoStop}%
\bibitem [{\citenamefont {Angstmann}\ \emph {et~al.}(2004)\citenamefont
  {Angstmann}, \citenamefont {Dzuba},\ and\ \citenamefont
  {Flambaum}}]{AngDzuFla04}%
  \BibitemOpen
  \bibfield  {author} {\bibinfo {author} {\bibfnamefont {E.~J.}\ \bibnamefont
  {Angstmann}}, \bibinfo {author} {\bibfnamefont {V.~A.}\ \bibnamefont
  {Dzuba}}, \ and\ \bibinfo {author} {\bibfnamefont {V.~V.}\ \bibnamefont
  {Flambaum}},\ }\href {\doibase 10.1103/PhysRevA.70.014102} {\bibfield
  {journal} {\bibinfo  {journal} {Phys. Rev. A}\ }\textbf {\bibinfo {volume}
  {70}},\ \bibinfo {pages} {014102} (\bibinfo {year} {2004})}\BibitemShut
  {NoStop}%
\bibitem [{\citenamefont {Flambaum}\ and\ \citenamefont {Kozlov}()}]{FlaKoz08}%
  \BibitemOpen
  \bibfield  {author} {\bibinfo {author} {\bibfnamefont {V.~V.}\ \bibnamefont
  {Flambaum}}\ and\ \bibinfo {author} {\bibfnamefont {M.~G.}\ \bibnamefont
  {Kozlov}},\ }\href@noop {} {}\bibinfo {note} {{in {\it Cold molecules.
  Theory, experiment, applications}, edited by R. V. Krems, W. C. Stwalley, and
  B. Friedrich (CRC Press, Boca Raton, FL, 2009), p.~597; e-print
  arXiv:0711.4536v2}}\BibitemShut {NoStop}%
\bibitem [{Koz()}]{Kozlovprivate}%
  \BibitemOpen
  \href@noop {} {}\bibinfo {note} {{M. G. Kozlov (private
  communication)}}\BibitemShut {NoStop}%
\bibitem [{\citenamefont {Flambaum}\ and\ \citenamefont
  {Kozlov}(2007{\natexlab{b}})}]{FlaKoz07a}%
  \BibitemOpen
  \bibfield  {author} {\bibinfo {author} {\bibfnamefont {V.~V.}\ \bibnamefont
  {Flambaum}}\ and\ \bibinfo {author} {\bibfnamefont {M.~G.}\ \bibnamefont
  {Kozlov}},\ }\href {\doibase 10.1103/PhysRevLett.99.150801} {\bibfield
  {journal} {\bibinfo  {journal} {Phys. Rev. Lett.}\ }\textbf {\bibinfo
  {volume} {99}},\ \bibinfo {pages} {150801} (\bibinfo {year}
  {2007}{\natexlab{b}})}\BibitemShut {NoStop}%
\end{thebibliography}

%

\end{document}